\documentclass[preprint2]{aastex61}

\usepackage{graphicx}
\usepackage{epsfig}
\usepackage{times}
\usepackage{natbib}
\usepackage{url}
\usepackage{color}
\usepackage{amssymb,amsmath}
\usepackage{gensymb}
\usepackage{comment}

\hypersetup{linkcolor=red,citecolor=cyan,filecolor=cyan,urlcolor=blue}

\newcommand{\eg}{e.g.}

\begin{document}

\title{Sequential eruptions triggered by flux emergence - observations and modeling}

\author[0000-0001-7572-2903]{S.~Dacie}
\affiliation{Mullard Space Science Laboratory, University College London, Holmbury St. Mary, Surrey, RH5 6NT, UK}
\author[0000-0003-3843-3242]{T.~T\"or\"ok}
\affiliation{Predictive Science Inc., 9990 Mesa Rim Road, Suite 170, San Diego, CA 92121, USA}
\author[0000-0001-8215-6532]{P.~D\'emoulin}
\affiliation{Observatoire de Paris, LESIA, UMR 8109 (CNRS), F-92195 Meudon-Principal Cedex, France}
\author[0000-0002-4459-7510]{M.~G.~Linton}
\affiliation{Naval Research Laboratory, Washington, DC 20375, USA}
\author[0000-0003-1759-4354]{C.~Downs}
\affiliation{Predictive Science Inc., 9990 Mesa Rim Road, Suite 170, San Diego, CA 92121, USA}
\author[0000-0002-2943-5978]{L.~van Driel-Gesztelyi}
\affiliation{Mullard Space Science Laboratory, University College London, Holmbury St. Mary, Surrey, RH5 6NT, UK}
\affiliation{Observatoire de Paris, LESIA, UMR 8109 (CNRS), F-92195 Meudon-Principal Cedex, France}
\affiliation{Konkoly Observatory of the Hungarian Academy of Sciences, Budapest, Hungary}
\author[0000-0003-3137-0277]{D.~M.~Long}
\affiliation{Mullard Space Science Laboratory, University College London, Holmbury St. Mary, Surrey, RH5 6NT, UK}
\author[0000-0003-0072-4634]{J.~E. Leake}
\affiliation{NASA Goddard Space Flight Center, 8800 Greenbelt Rd, Greenbelt, MD 22071, USA}
\correspondingauthor{Tibor T\"or\"ok}
\email{tibor@predsci.com}

\shorttitle{Eruptions triggered by flux emergence} 
\shortauthors{Dacie et al.}

\begin{abstract}
We describe and analyze observations by the {\em Solar Dynamics Observatory} of the emergence of a small, bipolar active region within an area of unipolar magnetic flux that was surrounded by a circular, quiescent filament. Within only eight hours of the start of the emergence, a partial splitting of the filament and two consecutive coronal mass ejections took place. We argue that all three dynamic events occurred as a result of particular magnetic-reconnection episodes between the emerging bipole and the pre-existing coronal magnetic field. In order to substantiate our interpretation, we consider three-dimensional magnetohydrodynamic simulations that model the emergence of magnetic flux in the vicinity of a large-scale coronal flux rope. The simulations qualitatively reproduce most of the reconnection episodes suggested by the observations; as well as the filament-splitting, the first eruption, and the formation of sheared/twisted fields that may have played a role in the second eruption. Our results suggest that the position of emerging flux with respect to the background magnetic configuration is a crucial factor for the resulting evolution, while previous results suggest that parameters such as the orientation or the amount of emerging flux are important as well. This poses a challenge for predicting the onset of eruptions that are triggered by flux emergence, and it calls for a detailed survey of the relevant parameter space by means of numerical simulations.
\end{abstract}


%%%%%%%%%%%%%%%%%%%%%%%%%%%%%%%%%%%%%%%%%%%
%%%%%%%%%%%%%%%%%%%%%%%%%%%%%%%%%%%%%%%%%%%
\section{Introduction}
\label{s:int}
%%%%%%%%%%%%%%%%%%%%%%%%%%%%%%%%%%%%%%%%%%%
%%%%%%%%%%%%%%%%%%%%%%%%%%%%%%%%%%%%%%%%%%%
%
Coronal mass ejections (CMEs) are eruptions of sheared or twisted magnetic fields from the solar corona. They may contain a filament consisting of dense and cool material initially suspended above the solar surface. Many mechanisms have been proposed to explain how CMEs (and associated flares) are initiated, causing the magnetic structure to rise \citep[see, e.g.,][]{aulanier14}.  At a critical height, the structure is believed to become torus unstable \citep{Kliem06}, which causes it to rapidly accelerate upward. At the same time, a current sheet forms between oppositely orientated field lines beneath the unstable structure. Reconnection within the current sheet leads to flaring \citep[as in the CSHKP model][]{Carmichael64, Sturrock66, Hirayama74, Kopp76} and additional acceleration of the ejecta.

One possible mechanism for the initiation of eruptions is flux emergence nearby a preexisting, current-carrying magnetic structure, as described by, e.g., \cite{Chen00}. In their simulation labeled as case B, bipolar field emerges near to a flux rope that is in stable equilibrium with the ambient field. The new flux is orientated ``favorably'' for reconnection, as defined by \citet{Feynman95}, meaning that the orientation of the emerging bipole is chosen such that a current layer forms {\em between} the flux rope and the bipole. As reconnection occurs across the current layer, two new sets of field lines are created; a small arcade that connects the emerging flux and the ambient field, and long field lines that arch over the flux rope and connect to the other polarity of the emerging flux (see Figures\,5b in \citealt{Chen00} and 3a in \citealt{williams05}). These latter field lines become somewhat longer due to the displacement of one of their foot points, so their downward acting magnetic tension decreases. This ongoing reduction of magnetic tension leads to a continuous slow rise of the flux rope, which finally leads to loss of equilibrium (or torus instability) and the eruption of the rope. The eruption is potentially facilitated also by an overall expansion of the ambient field due to the changes in the photospheric flux distribution \citep{Ding08}.    

There have been many observational studies of newly emerging flux occurring before CMEs \citep{Feynman95, Wang99, Jing04, Xu08, schrijver09}. These have shown that the most favorable conditions for triggering a CME arise when the orientation of the emerging flux is opposite to that of the existing field and when the emergence occurs close to the polarity inversion line \citep[PIL; e.g.,][]{Xu08}. However, \citet{Louis15} associated a flare and CME with flux emergence that was neither of favorable orientation nor located close to the PIL, and other exceptions can be found in the studies mentioned above. Other observations have shown magnetic flux emergence apparently causing a filament to split \citep{Li15}. These contrasting observations suggest that the set of conditions required for emerging flux to initiate an eruption are not yet fully understood \citep[see also][]{Lin01}.

%===================================================
\begin{figure*}[t] % Figure 1
\centering
\includegraphics[width=0.821\textwidth]{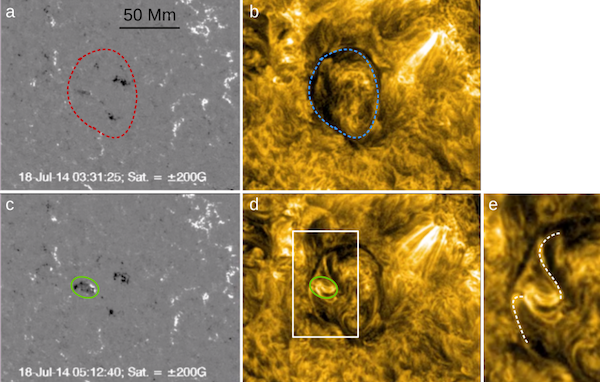}
\caption{
Apparent splitting of the eastern section of the filament. Length scales for panels (a)--(d) are indicated in (a). Shown are HMI magnetograms (left) and AIA\,171\,\AA\ images (right), at 03:30\,UT (top) and 05:12\,UT (bottom) on 18 July 2014. (a) Magnetogram and (b) quiescent filament around the onset of emergence. Red (blue) dashed lines outline the filament location. (c) Region of flux emergence (green oval), slightly to the west of the eastern filament section. (d) Interaction between emerging flux and pre-existing magnetic field. Bright streaks reach both northward and southward, suggesting new magnetic connectivities. (e) Zoom into the area shown as white rectangle in (d). White dashed lines outline the new connectivities. The AIA images shown here and in Figure\,\ref{f4} were processed using the Multi-scale Gaussian Normalization Technique of \citet{Morgan14}. An animation of this figure is available with the online version of this manuscript.
}
\label{f1}
\end{figure*}
%===================================================

Many modeling studies have been able to produce eruptions by introducing emerging flux into a pre-existing magnetic field configuration; either containing a potential or sheared arcade field \citep{Notoya07, Zuccarello08, Zuccarello09, Kusano12, jacobs12, roussev12, Kaneko14}, or a flux rope \citep[\eg][]{Chen00, Lin01, Xu05, Shiota05, Dubey06, Ding08}. Whether an eruption is produced depends on various parameters such as the strength of the new magnetic flux and its position and orientation. Changing these parameters can lead to cases where the emerging flux acts to additionally stabilize a flux rope rather than triggering its eruption \citep[\eg\ the cases shown in Figure\,7 of][]{Chen00}.

In this paper, we study the effects of the emergence of a small bipole nearby a quiescent circular filament on 18 July 2014. The eastern section of the filament is seen to partially split and to form new connectivities, followed by the eruption of its western section shortly after. A few hours later, a second eruption occurs above the PIL segment that has formed between the emerging flux and the pre-existing field, suggesting the formation of non-potential magnetic fields at this location during the emergence of the bipole. In Section\,\ref{s:obs}, we discuss the observations and propose mechanisms to explain these activities. In Section\,\ref{s:sim} we present magnetohydrodynamic (MHD) numerical simulations that qualitatively reproduce the filament splitting and the first eruption, and suggest a possible mechanism for the formation of a flux rope between the emerging and preexisting flux. Finally, we discuss the results and draw our conclusions in Section\,\ref{sect_Conclusions}.

%%%%%%%%%%%%%%%%%%%%%%%%%%%%%%%%%%%%%%%%%%%
%%%%%%%%%%%%%%%%%%%%%%%%%%%%%%%%%%%%%%%%%%%
\section{Observations and Proposed Mechanisms}
\label{s:obs}
%%%%%%%%%%%%%%%%%%%%%%%%%%%%%%%%%%%%%%%%%%%
%%%%%%%%%%%%%%%%%%%%%%%%%%%%%%%%%%%%%%%%%%%

%%%%%%%%%%%%%%%%%%%%%%%%%%%%%%%%%%%%%%%%%%%
\subsection{Data}
%%%%%%%%%%%%%%%%%%%%%%%%%%%%%%%%%%%%%%%%%%%
%
A quiescent circular filament and the newly forming active region NOAA 12119, which emerged within a negative polarity area encircled by the filament, close to its eastern section, were studied for the eight hours following the start of the active region's emergence at $\approx$\,03:30 UT on 18 July 2014 at [-376", -415"] in helioprojective-cartesian coordinates. The partial splitting of the filament and the two eruptions occurred during this time period. Data from the Atmospheric Imaging Assembly \citep[AIA;][]{Lemen12} on board the \textit{Solar Dynamics Observatory} \citep[SDO;][]{Pesnell12} were used to identify structures and connectivities in the corona, and photospheric line-of-sight magnetic field measurements, provided by the Helioseismic Magnetic Imager \citep[HMI;][]{Schou12} were used to calculate the magnetic flux of the emerging bipole. All images shown in Figs.\,\ref{f1} and \ref{f4} below were rotated to the observer's view at 05:55\,UT on July 18, which is roughly midway between the respective onset times of the partial filament splitting and the first eruption. The center of this field of view is at -23.7\degree\, longitude (in heliographic coordinates). We refrained from rotating all images to the central meridian, to minimize the interpolation of the data.

%===================================================
\begin{figure}[t] %Figure 2 
\centering
\includegraphics[width=0.4\textwidth]{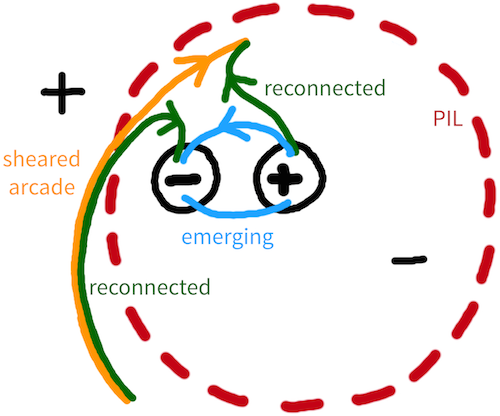}
\caption{
Diagram showing the splitting mechanism. The red dashed circle represents the PIL of the pre-emergence magnetic configuration, above which the filament resides (Figure\,\ref{f1}(b)). The emerging bipole is sketched with + and - signs in black circles, and blue field lines representing its magnetic connectivities. The orange line, which crosses the PIL, represents a highly sheared field line of the eastern filament arcade. It reconnects with the emerging bipole, forming the green field lines.
}
\label{f2}
\end{figure}
%===================================================

%%%%%%%%%%%%%%%%%%%%%%%%%%%%%%%%%%%%%%%%%%%
\subsection{Splitting of the Eastern Filament Section}
\label{ss:Obs_Splitting}
%%%%%%%%%%%%%%%%%%%%%%%%%%%%%%%%%%%%%%%%%%%
%
The quiescent filament (shown in Figure \ref{f1}(b) at 03:30 UT on 18 July 2014) was almost circular in shape and had formed between an area of dispersed negative field (inside the dashed line shown in Figure \ref{f1}(a),(b)) and positive field (outside). An emerging bipole, which later becomes active region NOAA 12119, began to emerge just to the west of the eastern section of the circular filament. The orientation of its magnetic fields is mostly West to East, although the presence of magnetic tongues leads to some deviation of the PIL orientation from that direction (Figure \ref{f1}(c)). Following \citet{Poisson16}, these tongues are interpreted as the contribution of the azimuthal field component of the emerging flux tube. They indicate a negative magnetic helicity, which is also suggested by the shear of the loops seen in the corona (Figure \ref{f1}(d)).\footnote{\citet{Luoni11} showed that the helicity sign derived from a magnetic-tongue pattern agrees with other proxies, such as loop shear.}

As the new flux started to emerge, it immediately began to interact with the surrounding magnetic field, as is apparent from the formation of bright loops in the AIA 171~\AA\ images as early as 05:12 UT. These loops are shown in Figure\,\ref{f1}(d) and outlined by dashed white curves in the zoom shown in panel Figure\,\ref{f1}(e). They are indicative of new magnetic connectivities as a result of reconnection between the magnetic field of the emerging bipole and the magnetic structure supporting the filament, which is likely a highly sheared arcade.

Figure \ref{f2} is a top-down diagram showing the field lines of the sheared arcade before reconnection (orange), those of the emerging bipole (blue), and those formed by reconnection (dark green), which are of the same shape as the bright loops outlined in Figure \ref{f1}(e). This reconnection likely causes the sheared arcade (and thus the filament) to split, at least partially, with some of its flux becoming connected to the emerging bipole. 

\begin{figure}[t] %Figure 3
\centering
\includegraphics[width=0.392\textwidth]{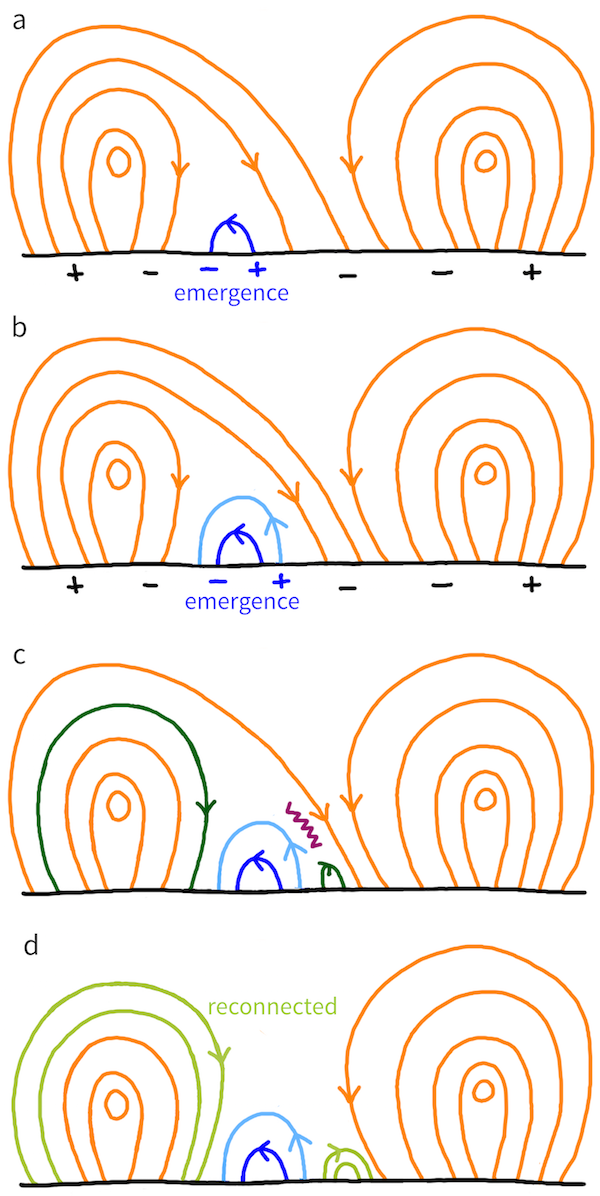}
\caption{
2D diagram showing the reconnection described in Section\,\ref{ss:Obs_Splitting}. The black line indicates the photosphere. Dark (light) blue field lines represent newly emerging flux (emerged in the previous panel). Orange field lines show a cross-section of the pre-existing field configuration; with the eastern (western) filament arcade on the left (right). Dark (light) green field lines are formed by reconnection (reconnected in the previous panel). The purple zigzag line represents a current sheet where reconnection takes place. The reconnection produces two new field-line sets, anchored in the negative and positive polarity of the bipole, respectively (cf. Figure\,\ref{f2}). Note that new flux continues to emerge in panels (c) and (d), but is omitted for clarity.
\label{f3}
}
\end{figure}

A schematic side-on view of this phase of the evolution is shown in Figure \ref{f3}, emphasizing reconnection of the emerging bipole with the field surrounding the highly sheared filament-arcade core. Here again, the orange field lines depict the filament arcade, blue depict the emerging bipole and green those formed by reconnection. Since the field lines are drawn on a 2D plane, those of the emerging bipole and those surrounding the core of the eastern filament arcade appear to be oriented parallel to each other, but this is not the case in reality. 

The dark green field line on the left of Figure \ref{f3}(c) is equivalent to that on the left of Figure \ref{f2}. This field line has been shortened by the reconnection (see Figure\,\ref{f2}), which increases its magnetic tension and leads to an additional stabilization of the core field. The small dark green line of Figure \ref{f3}(c) is equivalent to the smaller reconnected line shown in Figure \ref{f2}. In what follows, we refer to the new magnetic connection associated with the latter line as the ``new arcade''. Since the new arcade forms by reconnection between the bipole and the original filament arcade, it likely contains a significant amount of shear/twist, which may have been required for powering the second eruption described below.

\begin{figure*} % Figure 4
\centering
\includegraphics[width=0.8\textwidth]{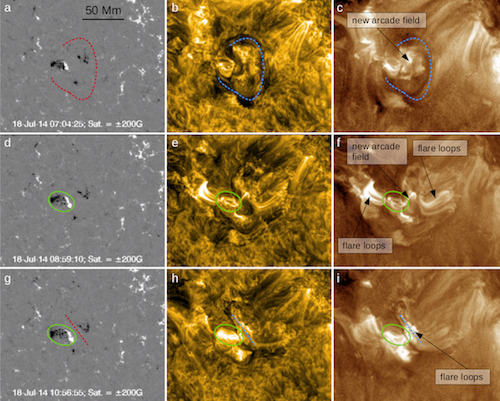}
\caption{
First and second eruption. The length scales for all panels is indicated in (a). Shown are HMI magnetograms (left), AIA\,171\,\AA\ (center) and 193\,\AA\ (right) images. (a)--(c) Before the first eruption, at 07:04 UT on 18 July 2014. Dashed lines indicate the location of the filament that undergoes the first eruption. The bright new arcade field lines in (c) are formed by reconnection of the emerging flux with the pre-existing filament arcade, as shown in Figure \ref{f5}(c). (d)--(f) Shortly after the first eruption, at 08:59 UT. The flare loops produced in the eruption can be seen in (e) and (f). The total unsigned magnetic flux of the emerging bipole was $\approx$\,4.7 $\times$ 10\textsuperscript{20}\,Mx, and it had a size of about $25 \times 12$\,Mm at this time. (g)--(i) Just after the second eruption, at 10:56 UT. The dashed lines indicate the PIL of the erupted arcade. Sheared flare loops produced in the eruption can be seen in (h) and (i). The flux-emergence region is outlined by green ovals in panels (d)--(i). At this time, the emerging bipole has reached a total size of about $28 \times 13$\,Mm and a total flux of $\approx$\,6.0 $\times$ 10\textsuperscript{20}\,Mx. An animation of this figure is available with the online version of this manuscript.
}
\label{f4}
\end{figure*}

Due to plasma heating caused by reconnection, it is difficult to follow the evolution of the filament material involved in this reconfiguration. It appears that some of it ended up in the new arcade, as the observations show the presence of a north-south directed, S-shaped filament section that seems to follow that structure (Figures\,\ref{f1}(e) and \ref{f4}), albeit some filament material may have been present at that location prior to the emergence (Figure\,\ref{f1}(b)).

%===================================================
%===================================================

%===================================================
%===================================================

%%%%%%%%%%%%%%%%%%%%%%%%%%%%%%%%%%%%%%%%%%%
\subsection{First Eruption}
\label{ss:Obs_Eruption1}
%%%%%%%%%%%%%%%%%%%%%%%%%%%%%%%%%%%%%%%%%%%
%
The western part of the filament is seen to start rising slowly at $\approx$\,07:00 UT. Around 07:45 UT, the rise accelerates and the western part of the filament fully lifts off. It erupts strongly non-radially  eastward, over the eastern part of the filament, and seems to drag the latter with it. It thus appears that the whole filament erupts (Figure \ref{f4}(e)), except perhaps those sections that were disconnected during the earlier phase of the bipole emergence. The flare loops produced by this eruption can be clearly seen in AIA 171 and 193~\AA\ images, as shown in Figure \ref{f4}(e),(f). The CME associated with this eruption is first seen in data from the Large Angle and Spectrometric Coronagraph \citep[LASCO;][]{Brueckner95} C2 on board the \textit{Solar and Heliospheric Observatory} (SOHO) at $\approx$\,09:35.

\begin{figure*} % Figure 5
\centering
\includegraphics[width=1.\textwidth]{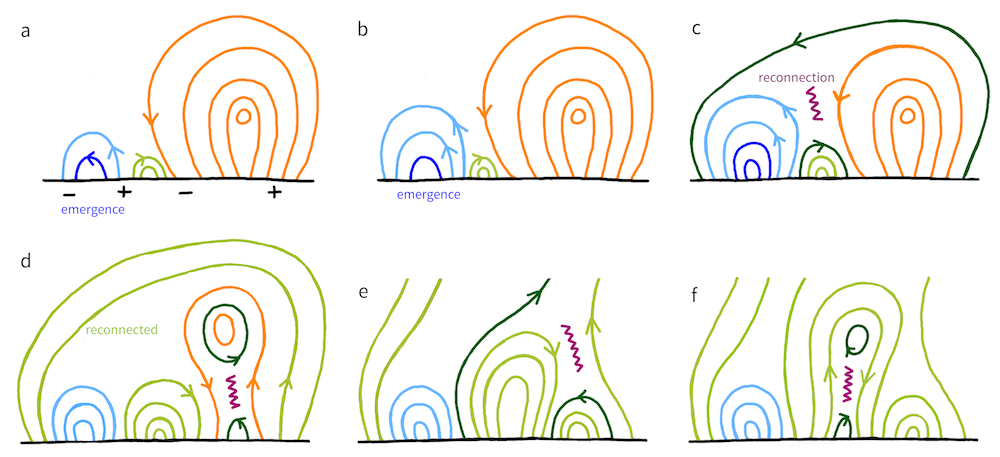}
\caption{
Same as Figure\,\ref{f3}, now for the mechanisms believed to produce the two eruptions. (a) Emerging bipole (blue), {\bf new arcade} (green), and western filament-arcade (orange). The eastern filament-arcade does not participate in the evolution and is omitted here. (b) Continuing bipole emergence and expansion in the corona. (c) Reconnection between bipole and western filament-arcade, forming long overlying field lines and smaller loops above the new arcade. The lengthening of the overlying field lines reduces the magnetic tension on the western filament-arcade, allowing it to expand. (d) Eruption of the western filament-arcade and associated reconnection underneath the filament, cutting its ties to the photosphere and further accelerating it upwards. (e) Expansion of the new arcade induces reconnection with locally open field lines left behind from the eruption, accelerating its rise. (f) Second eruption and flare loops formed by the reconnection in the wake of the eruption. 
\label{f5}
}
\end{figure*}

The observations are interpreted as shown in Figure \ref{f5}(a)--(d) and described as follows. After the emerging bipole has `eaten through' all of the field lines of the eastern arcade, it can start to reconnect with the western arcade. This reconnection (Figure \ref{f5}(c)) produces two new sets of field lines (dark green); small loops and long overlying ones. 

The bright loops seen in Figure \ref{f4}(c) before the eruption are interpreted as these small loops, which connect the positive polarity of the emerging bipole and the negative polarity of the filament arcade. They are expected to accumulate above the sheared new arcade that formed earlier on, during the reconnection between the emerging bipole and the eastern filament arcade (Figures\,\ref{f2} and \ref{f3}(c),(d)). 

The long overlying loops produced by the reconnection shown in Figure\,\ref{f5}(c) have a lower magnetic tension than the field lines that were overlying the filament originally, allowing the western filament arcade to rise. At some point of the evolution the magnetic configuration becomes unstable, possibly due to loss of equilibrium (or torus instability) and erupts. Reconnection beneath the filament (Figure \ref{f5}(d)) produces the western flare loops shown in Figure \ref{f4}(f). This is the same process as described in case B of \citet{Chen00} for an eruption caused by flux emergence nearby a flux rope, and as demonstrated for a fully three-dimensional (3D) configuration in Section\,\ref{ss:sim_erupt}. 

%===================================================
%===================================================

After the eruption, flare loops can also be seen to the east of the emerging bipole (Figure\,\ref{f4}(f)). The observations indicate that the eruption of the western section of the filament likely destabilized the whole magnetic structure that overlies the PIL shown in dark red in Figure\,\ref{f4}(a), which appears reasonable since the highly sheared field carrying the filament was likely extending over the whole PIL. This means that the shortening of the field lines shown in Figure\,\ref{f3}(c),(d) was not sufficient to stabilize the configuration, which again appears reasonable as those field lines were relatively large. In this scenario, the eruption is expected to form loops along the entire length of the PIL, but not all of these are observed, probably due to differences in the plasma density associated with the local amount of reconnected flux (less energy is liberated in weaker magnetic fields). As for the reconnection described in Section\,\ref{ss:Obs_Splitting}, this 3D effect along the circular PIL cannot be depicted in the 2D cartoons of Figure \ref{f5}, which represents only a 2D cut of the configuration on the western side of the emerging bipole.

%%%%%%%%%%%%%%%%%%%%%%%%%%%%%%%%%%%%%%%%%%%
\subsection{Second Eruption}
\label{ss:Obs_Eruption2}
%%%%%%%%%%%%%%%%%%%%%%%%%%%%%%%%%%%%%%%%%%%
%
The second eruption originates above the new PIL between the positive polarity of the emerging bipole and a pre-existing negative flux concentration (dashed red line in Figure \ref{f4}(g)). The eruption occurs just a few hours after the first eruption, at $\approx$\,10:30 UT. Bright flare loops are seen after this second eruption (Figure \ref{f4}(h),(i)), showing the relaxation of a highly sheared arcade over a period of about 25 minutes. The CME associated with the second eruption enters the LASCO C2 field of view at $\approx$\,11:50 UT. It appears to travel faster than the first eruption, which may be due to a removal of some of the overlying coronal field by the first eruption.

The magnetic structure that most likely powers the second eruption is the new arcade that was formed by the reconnection process described in Section\,\ref{ss:Obs_Splitting}. It is indicated by the small green field lines on the right-hand side of the emerging bipole in Figure\,\ref{f3}(d). Magnetic flux is added to this new arcade during the reconnection that triggers the first eruption (Figure \ref{f5}(c)). The continuous westward motion of the leading positive polarity of the bipole towards the PIL of the new arcade likely concentrated the arcade's shear. Additionally, a highly twisted flux rope may have formed beneath the arcade by the process described in Section\,\ref{ss:sim_rope}. 

How is the second eruption initiated? As shown in Figure\,\ref{f5}(d), the first eruption leaves behind a region of reduced magnetic pressure, into which the sheared new arcade (or flux rope) can expand. This induces reconnection between the arcade and the erupting flux to its right-hand side (Figure\,\ref{f5}(e)). Note that this reconnection works in the opposite direction as the earlier one shown in Figure\,\ref{f5}(c): rather than adding closed flux to the arcade, it opens up flux on its top, thereby reducing the magnetic tension that holds down the arcade's sheared/twisted core. Such behavior has been previously observed, with flare ribbons moving backwards well after a CME was launched: see Figures\,11 and 12 in \cite{Goff07} for a similar reversal of the reconnection direction after the launch of a CME. This eventually facilitates the eruption of the core flux, which evolves into the second CME. Behind the eruption the reconnection shown in Figure \ref{f5}(f) is induced, which creates the flare loops seen in Figure \ref{f4}(h),(i). 

We note that the mechanism described here for the triggering of a second eruption due to a reduction of magnetic tension by a preceding eruption that occurs in an adjacent flux system is basically the same as modeled for ``sympathetic'' eruptions  by \cite{Torok11} and \cite{lynch13}; see also \cite{Gary04,devore05a,joshi.n16,li.s17}.

%%%%%%%%%%%%%%%%%%%%%%%%%%%%%%%%%%%%%%%%%%%
%%%%%%%%%%%%%%%%%%%%%%%%%%%%%%%%%%%%%%%%%%%
\section{Numerical Modeling}
\label{s:sim}
%%%%%%%%%%%%%%%%%%%%%%%%%%%%%%%%%%%%%%%%%%%
%%%%%%%%%%%%%%%%%%%%%%%%%%%%%%%%%%%%%%%%%%%
%
In this section we compare our interpretations of the observations with MHD simulations of the emergence of a strong and compact bipole in the vicinity of a large coronal flux rope. The simulations we consider here are part of a parametric study that was performed to study the triggering of CMEs by flux emergence \citep[as observed and analyzed by, e.g.,][]{Feynman95}. This study will be described in a forthcoming publication (T\"or\"ok et al., in preparation); here we restrict ourselves to a brief description of the basic setup. 

We emphasize that the simulations of our parametric study were not designed to reproduce the event analyzed in Section\,\ref{s:obs}, which results in a number of differences between the simulations and the observations (see below). Specifically, we are not intending here to reproduce the whole chain of the observed dynamic events (filament splitting, first and second eruption) in a single simulation. Rather, we choose from our parametric study three independent simulations that start from the same initial state and differ only in the distance between the pre-existing flux rope and the emerging bipole. Each simulation addresses only one of the observed dynamic events. Also, it should be kept in mind that the simulations use idealized configurations, i.e., they are not intended for a quantitative comparison with the observations. Instead, they should be considered merely as ``proof-of-concept'', serving to support our interpretations of the observations in terms of different reconnection processes and the resulting dynamics and system reconfigurations. We leave the design of a more realistic simulation of the observed events to a later investigation.

The simulations described here were performed using the MAS (Magnetohydrodynamic Algorithm outside a Sphere) code \citep[e.g.,][]{mikic94,lionello99}, which advances the standard viscous and resistive MHD equations. The $\beta=0$ approximation, in which thermal pressure and gravity are neglected, was used here, so that the evolution is driven by the Lorentz force. The use of this approximation is justified here, since the dynamics relevant for our investigation occur in corona, where the plasma beta is low. The spherical simulation domain covers the corona within 1.0--3.5 $R_\odot$, where $R_\odot$ is the solar radius. We note that, even though the lower boundary of the MAS domain is associated with the solar surface ($r=R_\odot$), it should physically be considered here as the bottom of the corona, since we use the $\beta=0$ approximation.

\begin{figure*}[t] % Figure 6
\centering
\includegraphics[width=1.\textwidth]{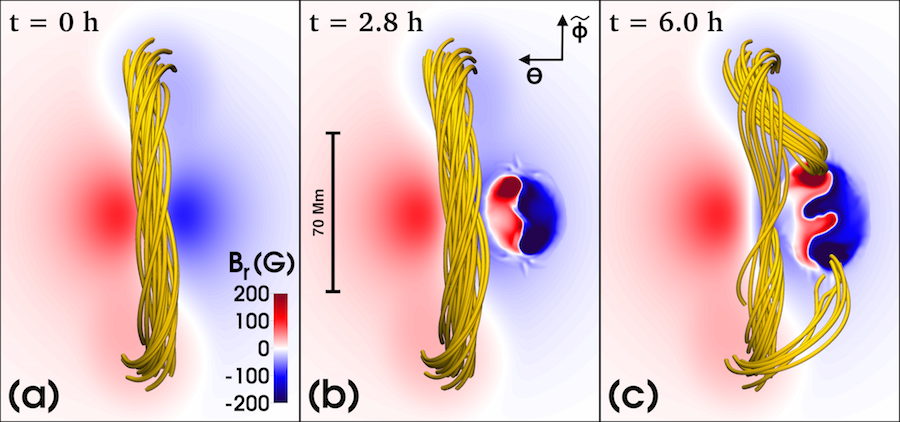}
\caption{
Simulation 1: Emergence of a bipolar flux region close to a pre-existing coronal flux rope. 
(a) Prior to emergence. Orange field lines depict the core of the TDm rope, which mimics 
the eastern section of the circular filament shown in Figure\,\ref{f1}.
(b) Early emergence. Emerging and TDm-background field lines and the current 
layer that forms between the emerging and pre-existing flux systems are omitted for clarity (see Figures\,\ref{f7} and \ref{f8}). 
(c) Later emergence. Reconnection between the emerging flux and the TDm 
rope has led to the formation of new connectivities similar to the observed ones (cf. Figure\,\ref{f1}(e)).
Length scales and coordinates are shown in (b).
}
\label{f6}
\end{figure*}

The initial coronal magnetic field consists of a flux rope embedded in a bipolar AR, as shown in a top-down view in Figure\,\ref{f6}(a). This configuration was constructed using the modified Titov-D\'emoulin model \citep[TDm;][]{titov14}, such that the flux rope is initially in stable magnetic equilibrium. The center of the TDm configuration is placed at the position $(r,\theta,\phi)=(1.,1.125,2.46)$, with $r$ in units of $R_\odot$ and $\theta$, $\phi$ in radians. The axis of the TDm flux rope is aligned with the $\phi$ axis. 
 
After relaxing the system until a sufficiently accurate numerical equilibrium is obtained, the emergence of a strong, compact bipolar AR is modeled ``kinematically'' (i.e., boundary-driven). To this end, horizontal slices of all three components of the magnetic field and the velocity are extracted at regular time intervals from an MHD simulation that used the Lare3D MHD code \citep{arber01} to model the emergence of a flux rope from the convection zone into a non-magnetized corona \citep{leake13}.\footnote{The simulation used here is very similar to the cases ``ND'' and ``ND1'' described in \cite{leake13}.} The slices are extracted at a height of the Lare3D simulation domain that corresponds approximately to the middle of the photosphere-chromosphere layer used in these simulations \citep[see][]{leake13}. The velocity components are directly imposed at the lower boundary of the MAS domain and used for the momentum equation in MAS. The radial magnetic field, $B_r(t)$, of the Lare3D simulation is superimposed for all slices on $B_r(r=R_\odot)$ of the TDm configuration. This superimposed component and the extracted tangential fields and velocities are then used to calculate the electric fields required for the induction equation in MAS (see \citealt{lionello13a} for details).    

An extensive parametric study of the resulting evolution was performed by varying the strength, location, and magnetic orientation of the emerging flux. Changing these parameters can change the interaction between the existing and emerging flux system. This leads in some cases to an eruption of the TDm flux rope (for similar studies see, e.g., \citealt{Chen00} and \citealt{Kusano12}).

%===================================================
%===================================================

In the simulations presented here, the bipolar AR emerges for about 1.5 hours at an almost constant rate of $\approx 5 \times 10^{20}$\,Mx\,h$^{-1}$, after which the emergence gradually slows down. After 6 hours, when the emergence has essentially saturated, the total unsigned flux of the AR is $\approx 1.3 \times 10^{21}$\,Mx, which is about 20 per cent of the total flux of the TDm configuration. At this time, the modeled bipole has reached a size of $\approx$\,50\,Mm (see Figure\,\ref{f6}(c)). 

We note that in our simulations the orientation of the polarity centers changes in the course of the emergence from east-west to north-south, which can be best seen by comparing panels (a) and (b) of Figure\,\ref{f8}. This was not the case for the real bipole, which essentially maintained an east-west orientation throughout the whole observed evolution. This indicates that the twist of the simulated emerging flux is larger than the twist of the real one. We believe that this difference does no affect the essential nature of the reconnection processes described in this section.

%===================================================
\begin{figure*}[t] % Figure 7
\centering
\includegraphics[width=1.\textwidth]{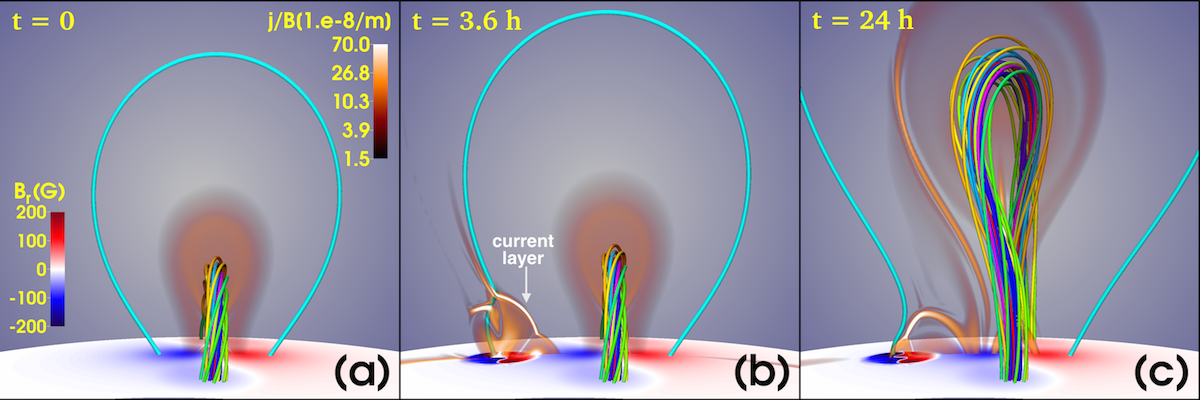}
\caption{
Simulation 2: Flux-rope eruption triggered by emerging flux. Here the TDm flux rope mimics the western section of the circular filament-arcade, so that in this view $\tilde{\phi}$ points towards the viewer and $\theta$ to the right (cf. Figure\,\ref{f6}). Shown are the core of the TDm rope (rainbow-colored field lines), an overlying field line (cyan), and electric currents in a transparent vertical plane perpendicular to the TDm rope axis (shown by $|{\bf j}|/|{\bf B}|$ in orange-white colors). (a) Initial configuration. The overlying field line is calculated starting from the positive (red) polarity of the TDm background field. (b) 3.6\,h later, after a substantial amount of flux has emerged. The overlying field has expanded and its negative (blue) footpoint has been displaced by reconnection in the current layer that separates the emerging flux from the background flux. (c) Configuration after 24\,h, showing the TDm flux rope in the process of eruption.
}
\label{f7}
\end{figure*}
%===================================================

The polarity signs of the magnetic configuration and the handedness of the TDm flux-rope were chosen in the parametric study without knowledge of the observed event described in this paper, and it turned out that they are opposite to those observed. Thus, when preparing the simulation data for the visualizations shown in Figures\,\ref{f6}--\ref{f8}, we generated an inverted coordinate, $\tilde\phi$ (see Figure\,\ref{f6}(a)), by mirroring the $\phi$ coordinate about $\phi=2.46$ (the center of the TDm configuration). The $\phi$-mirroring transforms the magnetic field from $[B_r(\phi),B_\theta(\phi),B_\phi(\phi)]$ to $[(B_r(\tilde\phi),B_\theta(\tilde\phi),-B_\phi(\tilde\phi)]$, keeping $\nabla \cdot {\bf B}=0$ and reversing only the $\phi$ component of the Lorentz force. This transformation changes the handedness of the TDm flux rope and of the emerging flux from negative to positive and from positive to negative, respectively, in agreement with the observations. We finally reverse the sign of ${\bf B}$, in order to reproduce the signs of all observed polarities. A corresponding procedure was applied to the current density, ${\bf j}$, which is used in Figures\,\ref{f7} and \ref{f8}. These transformations do not affect the evolution of the system, but significantly ease the visual comparison of the simulation results with the observations.

%%%%%%%%%%%%%%%%%%%%%%%%%%%%%%%%%%%%%%%%%%%
\subsection{Splitting of the TDm Flux Rope}
\label{ss:sim_split}
%%%%%%%%%%%%%%%%%%%%%%%%%%%%%%%%%%%%%%%%%%%
%
We first consider the simulation shown in Figure\,\ref{f6}, which we call ``simulation 1'' for further reference. In this run, the bipole emerges centered around $(r,\theta,\phi)=(1.,1.08,2.46)$, close to the TDm flux rope (at a distance of $0.045\,R_\odot$ in the $\theta$ direction), within the negative polarity of the TDm background field. This qualitatively corresponds to the situation shown in Figure\,\ref{f1}, namely to the emergence of the bipole close to the eastern section of the circular filament. The orientation of the emerging flux in the simulation is such that the initial axial-field direction of the emerging flux rope is antiparallel to the axial field at the core of the TDm flux rope.

\begin{figure*}[t] % Figure 8
\centering
\includegraphics[width=1.\textwidth]{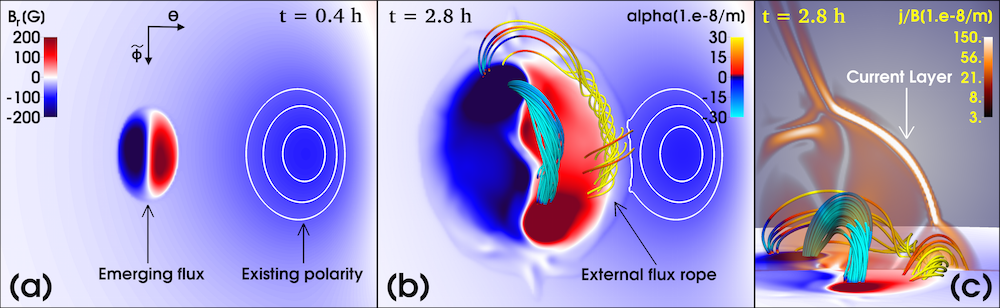}
\caption{
Simulation 3: emergence of a bipolar flux region close to a pre-existing flux concentration (similar to the observations shown in Figure\,\ref{f4}). The orientation of the configuration is the same as in Figure\,\ref{f7}; only the area containing the negative polarity of the TDm background field is shown. For better visibility, this polarity is highlighted by contour lines of -70, -80, and -90\,G in panels (a) and (b). (a) Early phase of emergence. (b) 2.4 hours later. Field lines connecting the emerging-polarity centers outline the core of the emerging flux rope (cyan). A new flux rope has formed at the external PIL between the bipole and the pre-existing flux concentration (yellow). Field lines are colored by the force-free parameter $\alpha={\bf j}\cdot{\bf B}/B^2$; note that the two ropes have opposite handedness. Some arcade-like field lines (with smaller $|\alpha|$) are shown above the new flux rope. (c) Tilted view of panel (b) along the PIL, showing additionally $|{\bf j}|/|{\bf B}|$ in a transparent vertical plane. 
}
\label{f8}
\end{figure*}

Due to the vicinity of the emerging flux to the TDm rope, the two flux systems start to interact early on in the evolution. Initially, only field lines of the potential field surrounding the TDm flux rope come into contact with the outer field lines of the emerging flux. Since the field direction of these flux systems is essentially antiparallel, a current layer similar to the ones shown in Figures\,\ref{f7} and \ref{f8} is formed between them. Driven by the expansion of the emerging flux in the corona, reconnection across this layer sets in. Once the outer flux regions have reconnected, the reconnection continues, now involving inner flux regions of the emerging bipole and the TDm flux rope.

Figure \ref{f6}(c) shows a situation at which a considerable fraction of the TDm flux rope has already reconnected to form new connections between the rope's foot points and the polarity centers of the emerging bipole. Being a result of reconnection, the corresponding field lines should appear bright in emission, just as the two streaks highlighted in Figure \ref{f1}(e). The morphological agreement between those streaks and the simulated new connectivities supports our interpretation that the emergence of the bipole resulted (at least partially) in the splitting of the flux rope or arcade that was carrying the eastern section of the circular filament (see Section\,\ref{ss:Obs_Splitting}).

Due to the initial north-south orientation of the emerging flux rope in the simulation, the field lines of the TDm flux-rope core and of the core of the emerging flux rope are oriented essentially antiparallel when they come into contact and reconnect (Figure\,\ref{f6}). This was not the case in the observed event, where the corresponding field directions were approximately perpendicular to each other. Such an orientation should, however, still allow a reconnection of the type shown in Figure\,\ref{f6}(c) to occur, as long as the interacting field lines are not close to being parallel \citep[e.g.,][]{linton01}. Indeed, in another simulation of our parametric study (not shown here), in which the orientation of the emerging flux rope was rotated by $3\,\pi/8$ (56\degree) clockwise compared to simulation 1, we still found strong reconnection between the bipole and the TDm rope and the development of new connectivities very similar to those shown in Figure\,\ref{f6}(c).

%%%%%%%%%%%%%%%%%%%%%%%%%%%%%%%%%%%%%%%%%%%
\subsection{First Eruption}
\label{ss:sim_erupt}
%%%%%%%%%%%%%%%%%%%%%%%%%%%%%%%%%%%%%%%%%%%
%
Our second simulation (simulation 2) is shown in Figure\,\ref{f7}. In this simulation, the TDm flux rope represents the western section of the circular filament. The orientation of the emerging bipole is the same as in simulation 1, but its center is now located at $(r,\theta,\phi)=(1.,1.04,2.46)$, about twice further away ($0.085\,R_\odot$ in the $\theta$ direction) from the TDm rope (just outside of the negative flux concentration of the TDm background field). The larger distance reflects the fact that in the observed case the western filament section was further away from the emerging bipole than the eastern section. The initial configuration of the simulation is shown in panel (a), where the cyan field line represents the potential background field overlying the TDm flux rope. As can be seen in panel (b), the emerging polarities and the TDm background polarities together form a quadrupolar polarity pattern, corresponding to what \cite{Feynman95} termed ``favorable for reconnection''. Note that the view in the figure is chosen such that the bipole emerges to the left (to the east) of the flux rope, as it was the case in the observations. 

As the new flux emerges, a current layer forms between the emerging flux and the TDm background field. Reconnection across this layer displaces field-line foot points of the background field from the edge of the negative TDm background polarity to the negative polarity of the emerging flux, i.e., further away from the TDm flux rope (Figure\,\ref{f7}(b)). The length of those field lines thus increases and they start to expand, which reduces the magnetic tension above the TDm rope. 

However, reconnection is not the only mechanism leading to such expansion. As numerically demonstrated by \cite{Ding08}, adding a small bipole to a 2D flux-rope configuration changes the configuration in such a way that the magnetic field overlying the flux rope is more expanded, as long as the bipole is placed close to the rope and in an orientation ``favorable for reconnection''. The expansion is merely due to the change in the boundary condition of the system \citep[see also][]{Wang99}; reconnection is not required. This effect takes place in our simulation, as the slowly emerging flux changes the boundary conditions of the coronal magnetic field. Due to the relatively large Alfv\'en speed in the corona, this information has sufficient time to travel into the domain and to affect the coronal magnetic field. 

The combined action of these two mechanisms is visualized in Figure\,\ref{f7}(b): the cyan field line has just reconnected with the emerging flux (see the strong kink of the field line at the position of the current layer), and its foot point on the left-hand side of the TDm rope has been displaced further away from the rope. Note that the field line has already expanded at the time it reconnects. This is partly due to the changes at the boundary, and partly due to the fact that field lines above it have reconnected and expanded earlier in the evolution.     

%===================================================
%===================================================

As a result of the continuous weakening of the magnetic tension due to field-line expansion, the TDm flux rope eventually cannot be stabilized anymore and erupts (Figure\,\ref{f7}(c)). The top of the rope rotates clockwise (when viewed from above), due to its right-handed twist \citep[e.g.,][]{green07,torok10}. Note that the initial opposite axial-field directions of the emerging flux rope and the TDm rope do not fundamentally affect the evolution in this case, since the TDm rope starts to erupt before it would significantly reconnect with the emerging flux. 

In the simulation, the eruption sets in about one day after the beginning of the flux emergence, which is much later than in the real event, where the time difference was about four hours. The onset time of the eruption depends on various parameters, predominantly on ``how far'' the TDm rope is initially from an unstable state, and how efficiently the emergence and associated reconnection act in weakening the stabilizing tension of the overlying flux. Changing the strength and position of the emerging flux and/or changing the initial current in the TDm rope will lead to different onset times.  

To summarize, simulation 2 demonstrates that the scenario illustrated in Figure\,\ref{f5}(a)--(d), and modeled in 2D for an infinitely long flux rope by \cite{Chen00}, can work also in fully 3D simulations, in which the foot points of the coronal flux rope are anchored in the photosphere. Thus, the simulation supports our interpretation put forward in Section\,\ref{ss:Obs_Eruption1}, namely that the first eruption was caused by a reduction of magnetic tension above the western part of the filament, as a result of flux emergence with an orientation ``favorable for reconnection''.

%%%%%%%%%%%%%%%%%%%%%%%%%%%%%%%%%%%%%%%%%%%
\subsection{Flux-Rope Formation Before Second Eruption}
\label{ss:sim_rope}
%%%%%%%%%%%%%%%%%%%%%%%%%%%%%%%%%%%%%%%%%%%
%
As described in Section\,\ref{ss:Obs_Eruption2}, the second eruption originates from the PIL that forms between the positive polarity of the emerging bipole and a neighboring, pre-existing negative flux concentration (Figure\,\ref{f4}(g)). In order for the eruption to occur, the flux residing above this PIL must have been non-potential. Since we found no indications for the presence of a PIL at this location prior to the emergence of the bipole, the corresponding shear/twist must have accumulated during the emergence process. 

We suggested in Section\,\ref{ss:Obs_Eruption2} that a new sheared arcade formed at this location by the reconnection process described in Section\,\ref{ss:Obs_Splitting}, and that the shear further concentrated due to the westward motion of the positive polarity. However, it is not clear whether sufficient shear to power an eruption can build up solely by this process. In this subsection we suggest an additional mechanism by which non-potential magnetic fields may have built up at this PIL. To this end, we consider a third simulation (simulation 3). Note that we focus here only on the formation of the pre-eruptive structure; we do not aim to model the eruption itself. 

In simulation 3, the orientation of the emerging flux is the same as in the previous simulations, i.e., the configuration is again ``favorable'' for reconnection. The emergence is now centered around $(r,\theta,\phi)=(1.,1.06,2.46)$, at an intermediate distance ($0.065\,R_\odot$) from the TDm rope (Figure\,\ref{f8}). This simulation is intended to mimic the emergence of the observed bipole east of the existing negative flux concentration, shown in the left panels of Figure\,\ref{f4}. We note that the formation of a new flux rope between the emerging and existing polarities, as described below, occurs in a very similar manner also in simulations 1 and 2. The reason why we use simulation 3 here to illustrate this process is that we had already analyzed that particular simulation regarding flux-rope formation prior to writing this article. 

As in simulations 1 and 2, the expansion of the emerging flux in the corona leads to the formation of a thin current layer (Figure\,\ref{f8}(c)). The layer develops above the ``external'' section of the PIL, i.e., the section that divides the positive emerging polarity and the pre-existing negative flux (Figure\,\ref{f8}(a)). A complex dynamic evolution involving different types of reconnection in the current layer and downward directed flows leads to the formation of a low-lying, highly twisted flux rope (Figure\,\ref{f8}(b)). Note that this ``external'' rope is right-handed, while the less twisted, thicker flux rope that connects the polarity centers of the emerging flux is left-handed (see \citealt{leake13} and references therein for the formation mechanism of this ``central'' flux rope).    

The accumulation of twisted field lines above the external section of the PIL eventually ceases. However, reconnection in the current layer still continues, now producing sheared, arcade-type field lines that accumulate above the external flux rope (Figure\,\ref{f8}(b),(c)). This ongoing reconnection corresponds exactly to the one sketched in Figure\,\ref{f3}(c),(d) and described in Section\,\ref{s:obs}.    

The external flux rope forms in our simulation due to reconnection across the current layer in the corona, so the formation mechanism should be robust with respect to the way in which the flux emergence into the corona is modeled (in our case via kinematic emergence). To check this, we have recently simulated an analogous situation using the Lare3D code, in which the flux emergence into a pre-existing coronal magnetic field is modeled dynamically, i.e., though the buoyant rise of a flux rope through the convection zone. We found the formation of an external flux rope also in this simulation, which will be described in a forthcoming publication.

The mechanisms that lead to the formation of the external rope, the dependence of its formation and properties on parameters such as the amount of twist of the emerging flux, and the implications of this structure for coronal jets and filaments that form between active regions will be discussed in detail in a forthcoming publication (T\"or\"ok et al., in preparation). For our purpose, the important point is that the development of such a flux rope during the emergence of new flux in the vicinity of a pre-existing polarity provides an additional explanation for the presence of highly non-potential magnetic fields along the PIL indicated by the dashed lines in Figures\,\ref{f4}(g)--(i).

%%%%%%%%%%%%%%%%%%%%%%%%%%%%%%%%%%%%%%%%%%%
%%%%%%%%%%%%%%%%%%%%%%%%%%%%%%%%%%%%%%%%%%%
\section{Summary and Conclusions}
\label{sect_Conclusions}
%%%%%%%%%%%%%%%%%%%%%%%%%%%%%%%%%%%%%%%%%%%
%%%%%%%%%%%%%%%%%%%%%%%%%%%%%%%%%%%%%%%%%%%
%
In this study, we investigated a magnetic configuration in which a filament resided above a circular PIL that encircled a dispersed negative polarity. We followed the early evolution of the small, bipolar active region NOAA 12119, which emerged within this polarity, close to the eastern section of the filament. Within eight hours of the onset of emergence, a partial splitting of the filament and two consecutive eruptions, both leading to CMEs, took place in the area. We utilized SDO data and MHD simulations to propose a scenario for the observed chain of events. 

Based on the observations, we propose that the bipole initially emerges completely within the arcade-field overlying the eastern section of the filament. Reconnection of the two flux systems leads to a shortening of the field lines surrounding the core field of the filament arcade and stabilizes the core field in this area (Figures\,\ref{f1}(d),(e), \ref{f2}, and \ref{f3}(c),(d)). This reconnection also causes at least a partial splitting of the field carrying the filament \citep[similar to the case of][]{Li15} and thereby produces a new arcade (and S-shaped filament) that connects the bipole with the original filament.

After the western side of the emerging bipole has reconnected through the eastern arcade, it starts to reconnect also with the field of the western arcade (Figure\,\ref{f5}(c)). This reconnection adds flux to the previously formed new arcade. Simultaneously, it destabilizes the western arcade, allowing the western section of the filament to rise and eventually erupt (Figures\, \ref{f4} and \ref{f5}(d)). 

Meanwhile, the continued emergence and westward motion of the leading polarity of the bipole may have concentrated the shear of the new arcade. Additionally, a highly twisted flux rope may have formed within it, as suggested by our simulation 3 (Section\,\ref{ss:sim_rope}). Since the first eruption has left behind a region of reduced magnetic pressure and weakened overlying field, the flux rope and surrounding new arcade can expand (Figure\,\ref{f5}(e)), eventually leading to the second eruption \citep[as in the sympathetic eruptions modeled by][]{Torok11}.

Our simulations support this scenario. In simulation 1 (Section\,\ref{ss:sim_split}), a bipolar flux region is emerged within one of the polarities of the TDm background field, close to the location of the pre-existing flux rope. The emerging and pre-existing TDm fields start to reconnect, and the TDm flux-rope field eventually forms new connectivities with the emerging bipole (Figure\,\ref{f6}(c)). The shapes and locations of these new connectivities correspond to the bright streaks seen in the observations (Figure\,\ref{f1}(d),(e)), suggesting that they were indeed a result of a partial splitting of the magnetic field carrying the eastern section of the filament. 

In simulation 2 (Section\,\ref{ss:sim_erupt}), the bipole is emerged further from the TDm flux-rope and with an orientation such that a quadrupolar polarity pattern is formed. This setup mimics the  interaction of the emerging flux with the western section of the filament. Both the changes in the boundary conditions caused by the emergence and reconnection between the two flux systems act to reduce the tension of the field overlying the TDm flux rope, which eventually leads to its eruption.  This provides an explanation for the first observed eruption, which begins in the western section of the circular filament, relatively far from the emerging bipole (Section\,\ref{ss:Obs_Eruption1}). 

In simulation 3 (Section\,\ref{ss:sim_rope}), we model the formation of a highly twisted flux rope over the PIL between one polarity of an emerging bipolar flux region and a pre-existing flux concentration of opposite polarity. This demonstrates that, in addition to the new arcade, also a highly twisted flux rope may have formed in the source region of the observed second eruption. This addition of non-potential magnetic field may make it easier to understand how the second eruption could originate in an area where concentrated sheared/twisted flux was not present prior to the flux emergence.      

We conclude that the position of the emerging bipole with respect to the background magnetic field configuration is a crucial factor for the interaction of these fields and the resulting evolution. Numerical simulations are able to qualitatively reproduce the various dynamic behavior observed for our case, and the upcoming study of T{\"o}r{\"o}k et al. will help to characterize the relationship between the position of emerging flux (and of other parameters such as its orientation, helicity sign, and amount of flux) and its interaction with the background field \citep[see also][]{Kusano12}. Even relatively small amounts of emerging flux may be able to trigger significant changes in the coronal field, increasing the difficulty to predict eruptions. More systematic observations and parametric numerical simulation studies would give us a better idea of the conditions under which it should be possible to predict coronal activity triggered by flux emergence.

\acknowledgments
We thank the referee for thoughtful and inspiring comments and Ron Moore for a helpful discussion regarding the second eruption. The authors are thankful to the SDO/ HMI and AIA consortia for the data. We also thank Z. Miki\'c for assisting in coupling the Lare3D simulations to MAS. S.D. acknowledges STFC for support via her studentship. L.v.D.G is partially funded under STFC consolidated grant number ST/N000722/1. L.v.D.G also acknowledges the Hungarian Research grant OTKA K-109276. D.M.L is an Early-Career Fellow funded by the Leverhulme Trust. T.T, C.D, J.E.L, and M.G.L acknowledge support from NASA's LWS and H-SR programs. M.G.L. acknowledges support from the Chief of Naval Research. We also thank the International Space Science Institute (ISSI) team 348 ``Decoding the Pre-eruptive Magnetic Configuration of Coronal Mass Ejections'' led by S. Patsourakos and A. Vourlidas.

\end{document}